# Polymer-free Assembly of Unencapsulated van der Waals Heterostructure Devices

*Han Xuan Wong,*[†] *Junjie Yu,*[†] *Yuyi Yan,* [†] *Adam Cronin,*[†] *Felix R. Fischer*[†,‡,§,ʃ] *

[†] Department of Chemistry, University of California, Berkeley, CA 94720, USA. [‡] Materials Sciences Division, Lawrence Berkeley National Laboratory, Berkeley, CA 94720, USA. [§] Kavli Energy NanoSciences Institute at the University of California Berkeley and the Lawrence Berkeley National Laboratory, Berkeley, California 94720, USA. [ʃ] Bakar Institute of Digital Materials for the Planet, Division of Computing, Data Science, and Society, University of California Berkeley; Berkeley, CA 94720, USA.

Van der Waals (vdW) heterostructures provide a versatile platform for exploring emergent quantum phenomena, yet their fabrication is often limited by interfacial contamination and incompatibility with surface-sensitive characterization. Here, we demonstrate muscovite mica as a polymer-free platform for assembling and patterning vdW heterostructure devices. Mica-mediated transfer enables clean exposed surfaces suitable for atomic-resolution scanning tunneling microscopy (STM) following mild thermal annealing, without aggressive post-processing. Mica also serves as a mechanically robust support for sequential pickup assembly, eliminating intermediate release steps. Furthermore, exfoliated mica flakes function as removable shadow masks for contact deposition, enabling straightforward patterning without conventional lithography. We quantify the temperature dependence of mica-mediated graphene pickup and demonstrate gate-dependent transport in a three-terminal graphene/hBN device fabricated using the approach. By integrating transfer, sequential assembly, and contact patterning within a single materials platform, mica provides a simple and accessible route to unencapsulated vdW heterostructures for surface-sensitive spectroscopy and quantum-device applications.

Van der Waals (vdW) heterostructures have emerged as a versatile platform for exploring correlated and topological quantum phases, including correlated insulating states and fractional Chern insulators, enabled by reduced dimensionality, enhanced Coulomb interactions, and tunable electronic band structures in atomically thin layers. As these emergent phases are highly sensitive to disorder, charge inhomogeneity, and structural imperfections, their realization and performance critically depend on maintaining pristine, atomically well-defined interfaces. Most commonly vdW heterostructures are assembled via sequential pick-up followed by encapsulation with inert layers,[1] leveraging the mutual adhesion between vdW materials. As an atomically flat, transparent insulator with few charged impurities, hexagonal boron nitride (hBN) is the ideal encapsulation material for transport measurements,[2] preventing direct contact between active layers and polymer-based transfer media that often introduce persistent residues.[3] Such devices enabled the landmark experimental observations of unconventional superconductivity[4] and correlated insulating phases[5] in prototypical magic angle twisted bilayer graphene (MATBLG). However, the full picture of these phenomena was only later resolved with the help of scanning tunneling spectroscopy (STS), which revealed the effect of local twist angle variation on correlation strength[6] and visualized nematic charge ordering and interaction-driven symmetry breaking in real space.[7] Beyond MATBLG, the unique capability of STM to probe the local electronic structure of low-dimensional systems with both atomic spatial and sub-meV energy resolution is especially relevant in the exploration of Wigner crystals[8] and charge density wave domain structures.[9]

For simple layered heterostructures, in situ growth from molecular precursors is often preferred for STM studies,[10, 11] yet a challenge arises when complex device architectures with twist angle control are required. STM is incompatible with typical fully encapsulated vdW heterostructures assemblies as it requires direct physical access to the active layer, and due to its extreme surface sensitivity, even monolayer surface contamination severely impacts measurements.[12] Extensive post-fabrication cleaning, including aggressive solvent treatments,[13] prolonged forming gas annealing,[7, 13, 14] or localized contact mode atomic force microscope (AFM) cleaning protocols[15-17] are thus required to access atomically clean surfaces in vdW heterostructures assembled using traditional polymer stamping. To streamline the preparation of clean, unencapsulated vdW heterostructures required for surface-sensitive techniques, recent transfer protocols based on silicon nitride and $MoS_2$ stamps have demonstrated polymer-free assembly, thereby significantly reducing surface contamination. However, silicon nitride stamps rely on complex microfabrication workflows,[18] while $MoS_2$ stamps suffer from limited optical transparency and variability in flake geometry,[19] losing out to conventional polymer stamps in terms of simplicity, repeatability, and ease of use.

Although it has long been known that muscovite mica exhibits strong adhesion to graphene,[20-24] it was only very recently that it has emerged as an alternative general transfer medium,[25] offering a polymer-free, highly transparent, and low-cost platform for deterministic assembly that overcomes the drawbacks of previous polymer-free stamps. Here, we show that muscovite mica enables an integrated, clean workflow for STM-compatible vdW device fabrication. *First*, mica-mediated transfer yields atomically clean surfaces that are directly compatible with STM requiring no additional cleaning steps beyond mild thermal annealing. *Second*, mica serves as both a transfer medium and a mechanically stable substrate, enabling sequential pick-up assembly without the need for a dedicated release step. *Third*, exfoliated mica flakes can be used to define clean, removable shadow masks via a pick-and-place process, eliminating the need for specialized shadow-mask deposition infrastructure for contact patterning. Taken together, these capabilities establish muscovite mica as a simple, inexpensive, and broadly accessible platform for end-to-end fabrication of vdW heterostructure device architectures fully compatible with the stringent requirements imposed by atomically resolved STM techniques.

Even under carefully controlled processing conditions, surfaces exposed to ambient atmosphere accumulate volatile hydrocarbons and aerosols,[26, 27] that degrade interfacial surface contact, adversely affect adhesion, and limit the reproducibility of transfer steps in vdW heterostructure assembly. Whereas non-layered polymer-free transfer stamps require thorough cleaning or surface treatment to reduce contamination,[18] pristine muscovite mica surfaces can readily be exfoliated by repeated mechanical cleavage of the surface layer of bulk crystals. We evaluated the cleanliness of the mica-mediated stamping process by transferring $MoS_2$ from $SiO_2$ to Au(111) and subsequently characterizing the transferred flakes by cryogenic ultra-high vacuum (UHV) STM. Au(111) was selected as the receiving substrate because freshly deposited or atomically cleaned Au(111) exhibits strong adhesion to transition metal dichalcogenides, while providing high electrical conductivity and atomically flat terraces well suited for STM measurements.[28] Figure 1a illustrates the transfer process, which was carried out under ambient conditions. Here, a bulk mica sheet (~4 cm$^2$, 10–20 μm thick) backed by a soft polydimethylsiloxane (PDMS) cushion (~0.1 cm$^2$) on a glass slide (Figure 1b) was aligned and gently pressed onto freshly cleaved $MoS_2$ flakes on $SiO_2$ (optical image in Figure 1c) at 120 °C for 10 min. After separation, the mica suspended above the PDMS cushion was decorated with multilayer $MoS_2$ flakes (optical image in Figure 1d). The stamp was then aligned and pressed onto a UHV cleaned Au(111) surface held at 120 °C for 5 min to release the $MoS_2$ flakes. The resulting Au(111) surface (optical image in Figure 1e) was transferred into a UHV preparation chamber, annealed for 2 h at 250 °C, and imaged using cryogenic scanning probe techniques. Figure 1f shows a typical large-scale (100 nm × 100 nm) topographic STM image recorded above a $MoS_2$ flake. The surface is atomically clean notwithstanding isolated particles unavoidable for surfaces briefly exposed to ambient conditions. Differential conductance point spectroscopy ($dI/dV$) recorded on the flake shows the expected band gap, $E_g \approx 1.4$ eV associated with multilayer $MoS_2$ (Figure 1g).[29, 30] High resolution STM images resolve the hexagonal lattice of S- and Mo-atoms lining the $MoS_2$ surface (Figure 1h). Notably, these high-quality topographic and spectroscopic data were acquired directly from $MoS_2$ surfaces that that had been in contact with the mica stamp. In contrast, untreated polymer-contacted surfaces are typically covered by multilayer residues that substantially degrade surface quality impacting imaging and spectroscopy in cryogenic UHV STM.[12] Our result shows that freshly cleaved mica stamps transfer 2D materials compatible with atomically resolved scanning probe microscopy (SPM) under ambient conditions, precluding the need for laborious post transfer processing and surface cleaning steps commonly associated with traditional polymer stamps. Additional point spectra at specific lattice sites and large-scale STM images on a separate transfer sample are included in Supporting Information Figure S1.

Beyond its superior cleanliness, the experiments above suggest that mica stamps may adhere to and release 2D materials through a mechanism distinct from that of conventional polymer stamps. Because mica is crystalline and does not undergo a glass transition, its adhesion is not expected to exhibit the strong temperature dependence characteristic of glassy polymers. Consistent with this picture, we qualitatively observed an increase in both pickup and release yields upon increasing the temperature from room temperature to 150 °C. The successful transfer of $MoS_2$ from $SiO_2$ to mica and subsequently to Au(111) may reflect a favorable hierarchy of interfacial energies, $\gamma_{MoS2\text{-}SiO2} < \gamma_{MoS2\text{-}mica} < \gamma_{MoS2\text{-}Au}$, with elevated temperature facilitating interfacial relaxation toward thermodynamic equilibrium. Mica is a layered ionic material with a more polar surface than $SiO_2$ and has been shown to interact strongly with charged and polarizable species, including proteins.[31-33] Such interactions may contribute to the strong adhesion of mica to polarizable 2D materials such as semiconductors and semimetals. The stronger interaction between $MoS_2$ and Au(111) can apparently overcome this adhesion, enabling release of $MoS_2$ onto the Au surface. In contrast, we were unable to release graphene from mica onto Au(111), highlighting that transfer success depends on the specific combination of 2D material and substrate. As discussed in the following section, however, when release is unfavorable, the mica stamp can instead serve directly as the substrate.

While the single-step transfer of $MoS_2$ onto Au(111) using muscovite mica stamps serves as a proof-of-concept, advanced device experiments compatible with SPM techniques often require multilayered, gated architectures. Most commonly a 2D material of interest is integrated into a capacitive device[6, 7] featuring a global back-gate used to tune the carrier density leaving the top surface exposed and accessible to SPM tools. Ideally both the gate metal and the dielectric should themselves be 2D materials to minimize strain and corrugation in the surface layer under investigation. Here we show that mica stamps can be used to assemble a high-quality multilayer graphene (MLG)/hBN/MLG heterostructures (Figure 2a). Following sequential pick-up of MLG (metallic back-gate), 32 nm hBN (gate dielectric), and 3 nm MLG (layer of interest) the bulk mica flake was detached from the glass/PDMS support and flipped to expose the trilayer vdW heterostructure shown in Figure 2b. Optical micrographs that document each pick-up step for multiple heterostructures, including those containing monolayer and bilayer graphene, are provided in Supporting information Figure S2–5.

In the following experiments we exclusively used graphene and hBN that had been exfoliated onto $O_2$ plasma-cleaned $SiO_2$ substrates. With mica stamps, we have found that graphene flakes with uniform thickness and lateral dimensions up to 1000 $\mu m^2$ could be picked up with > 90% yield (defined as either fully intact pickup or nearly complete pickup with only minor folds). The high yield was observed for contact temperatures of $T$ = 120–180 °C with a contact time of 5 min, providing a relatively broad processing window. At lower temperatures, the pickup yield decreased substantially (Figure 2h), which we tentatively attribute to slower removal of adsorbed species from the mica–graphene interface; a similar temperature dependence has been reported for silicon nitride stamps.[18] Higher temperatures were not investigated due to instrument limitations. Optical micrographs of the 37 flakes used to determine the temperature-dependent pickup yield are provided in Supporting Information Figures S6–S8. This behavior contrasts with AFM force–deflection measurements showing that the adhesion between mica and graphite decreases between room temperature and 130 °C,[25] suggesting that additional factors govern adhesion during an actual pickup process. We further found that nonuniform flake thickness and prolonged exposure to ambient conditions (> 1 h) substantially reduced the pickup yield.

For pickup of hBN using a graphene decorated mica stamp, we found that the highest yields were obtained for thin hBN flakes (<50 nm) whose contact area fully overlaps with the graphene. For thicker hBN, the graphene was occasionally released from the mica instead, whereas insufficient overlap could result in incomplete hBN pickup or folding of the non-overlapping regions. Conversely, when picking up graphene with hBN decorated mica stamps, the yield was essentially quantitative for few-layer graphene flakes of uniform thickness if the contact area fully overlapped with the hBN. Graphene regions extending beyond the hBN, however, were prone to folding during pickup. We also briefly explored releasing graphene and hBN onto freshly cleaned $SiO_2$, but found the process to be irreproducible, with most attempts resulting in partial release and flake tearing. These observations are consistent with our earlier inference that mica-mediated transfer is most reliable when the receiving substrate exhibits stronger adhesion to the transferred material than the donating substrate.

Figure 2b shows the fully assembled MLG/hBN/MLG/mica stack viewed from the top. AFM topographic images reveal uniform atomically flat surfaces largely free of contaminants (Figures 2c–e). Raman spectroscopy performed on a bilayer graphene (BLG)/hBN/FLG/mica heterostructure assembled the same way (Figure 2f,g) shows very sharp G-band features (FWHM = 14 $cm^{-1}$). The absence of the characteristic D-band commonly associated with defects suggests that the graphene layers remain undamaged in the transfer process. The polymer-free pick-up and flip assembly process outlined in Figure 2a is ideally suited for STM device fabrication. The approach leverages muscovite mica substrates as a mechanically robust and UHV compatible support that remains stable over a broad temperature range (0 K < $T$ < 1000 K) and exhibits negligible outgassing. The sequential pick-up and flip protocol obviates the need for a dedicated release step, reducing the strain cycles imposed on the top layer preserving the structural integrity of the surface accessible to SPM tools.

Reliable fabrication of electrical contacts remains a critical step in the preparation of vdW devices for SPM studies. While shadow-mask deposition represents an attractive alternative to conventional lithographic tools as it avoids exposing sensitive materials to polymers and etchants, it typically relies on custom microfabricated membrane masks and highly specialized alignment hardware.[12] Recently, masks based on transferred metal thin films have been demonstrated which have relaxed fabrication and alignment requirements,[34] however, they still require patterning by lithography or mechanical cutting with sub-μm sharp tips. We herein demonstrate an even simpler and more inexpensive masking strategy based on mechanically exfoliated muscovite mica flakes that can be implemented using the same transfer platform employed for device assembly (Figure 3a). Large-area mica flakes (> 0.01 $mm^2$) are readily obtained by exfoliation and can be manipulated with micrometer precision using a glass tip functionalized with a droplet of polypropylene carbonate (PPC) (Figures 3b–d), which functions as a thermal-release adhesive that can be volatilized with thermal treatment.[35] Importantly, PPC only locally contacts part of the top surface of the mica flakes and is far removed from the vdW heterostructure. Once positioned over selected regions of the device, the mica flakes serve as removable shadow masks during metal deposition, while larger areas of the substrate are protected with overlapping PDMS films. Following deposition of a 12 nm Ni layer, removal of the PDMS and mica masks revealed well-defined contacts without exposing the active device region to lithographic processing (Figure 3e). Beyond contact fabrication, the mica masks can be retained on the device until use, providing a convenient means of protecting air-sensitive surfaces during storage and handling. Detailed procedures for preparing the glass needles, manipulating mica flakes and positioning PDMS film pieces are including in Supporting Information Figure S9 and Note 1.

AFM characterization following mask removal revealed no detectable degradation of the device surface. The top MLG layer (MLG 1 in Figures 3f,g) is locally flat (RMS ~ 0.1 nm), uniform, and appears unperturbed when compared to the same region prior to metal deposition (Figures 2c,d). Figures 3g,h show sharp Ni contact edges, highlighting the intimate conformal contact established between the flexible mica mask and the substrate stack during deposition. The combination of conformal masking, contamination-free processing, and sharp pattern definition establishes exfoliated mica as a versatile platform for lithography-free contact fabrication in SPM compatible vdW device architectures.

To validate our stacking and contact patterning techniques, a 3-terminal single-layer graphene (SLG)/hBN/FLG field-effect transistor with 50 nm Pd contacts was fabricated (Supporting Information Figure S10–11). Transfer measurements (channel conductance $G$ against gate voltage $V_g$) showed the expected V-shaped curve for a graphene channel, with the charge neutrality point shifting nearer to zero bias after mild annealing at 170 °C, $1 \times 10^{-7}$ Torr for 12 h. This shift is expected for devices fabricated under ambient conditions and is attributed to the removal of adsorbed gases,[36] and the residual p-doping and electron-hole asymmetry may be explained by charge transfer from the Pd contacts.[37] This result shows that mica-mediated heterostructure assembly and contact patterning are compatible with established gated device architectures.

Our results establish muscovite mica as a versatile materials platform for transfer, support, and patterning in vdW device fabrication, providing a practical route toward cleaner surfaces and improved access to surface-sensitive measurements of low-dimensional quantum materials.

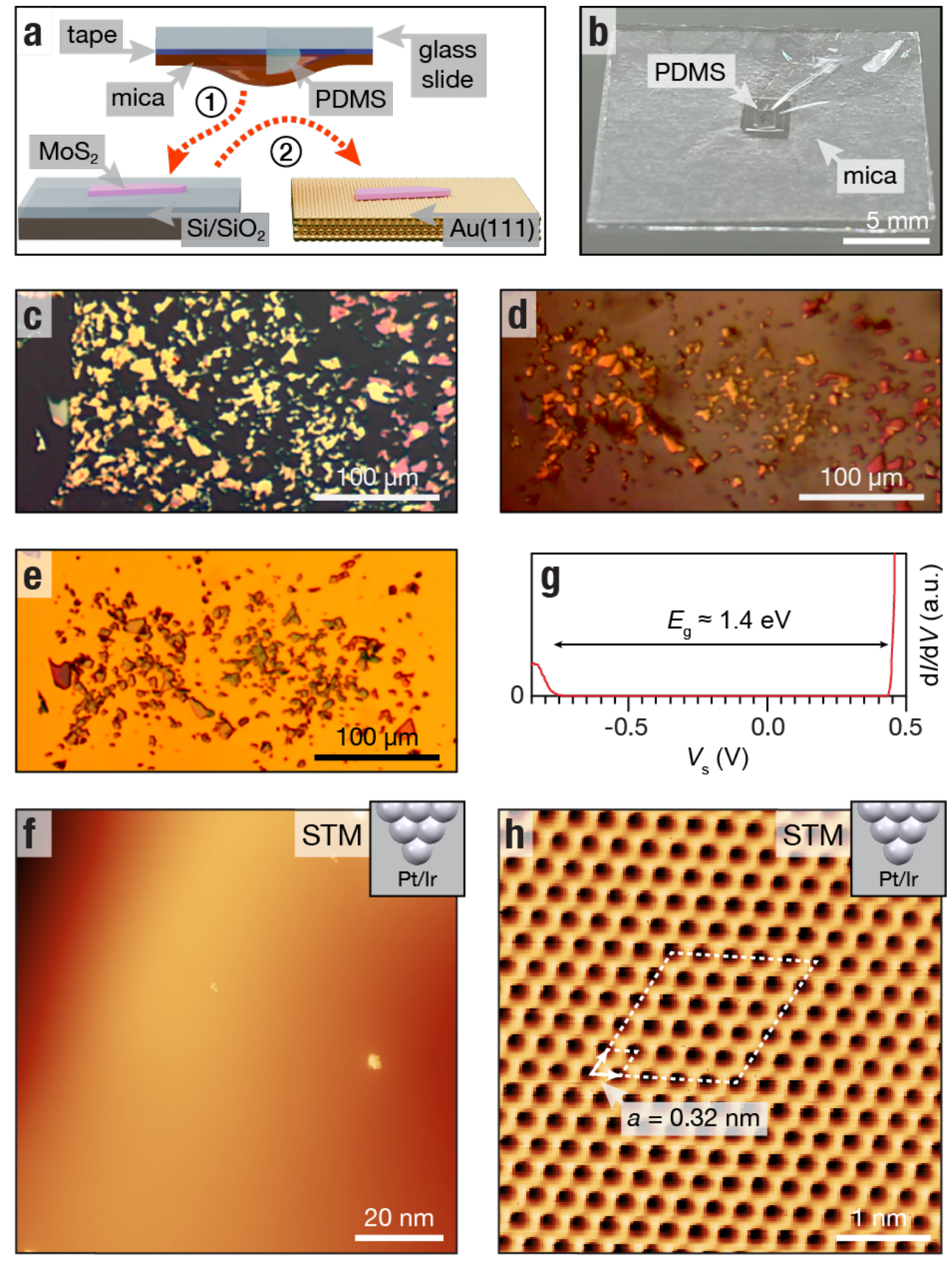

Figure 1. Pick-up and transfer of TMDs using mica stamps. (a) Schematic representation of the pick-up transfer steps of $MoS_2$ from $SiO_2$ to Au(111) using a mica stamp. (b) Image of PDMS-backed mica stamp (glass slide/PDMS/mica). (c) Optical micrograph of multilayer $MoS_2$ flakes exfoliated onto 100 nm $SiO_2$ on Si. (d) Optical micrograph of $MoS_2$ flakes picked up using a mica stamp. (e) Optical micrograph of $MoS_2$ flakes transferred onto Au(111). (f) Large-scale topographic STM image of transferred $MoS_2$ on Au(111) ($V_s$ = 1.0 V, $I_t$ = 5 pA). Bright features are residual adsorbates from exposure to ambient atmosphere. (g) d$I$/d$V$ spectrum obtained over $MoS_2$/Au(111) showing a band gap of $E_g \approx 1.4$ eV. (h) Atomic-resolution STM image of transferred $MoS_2$ on Au(111) ($V_s = 1.0$ V, $I_t = 400$ pA).

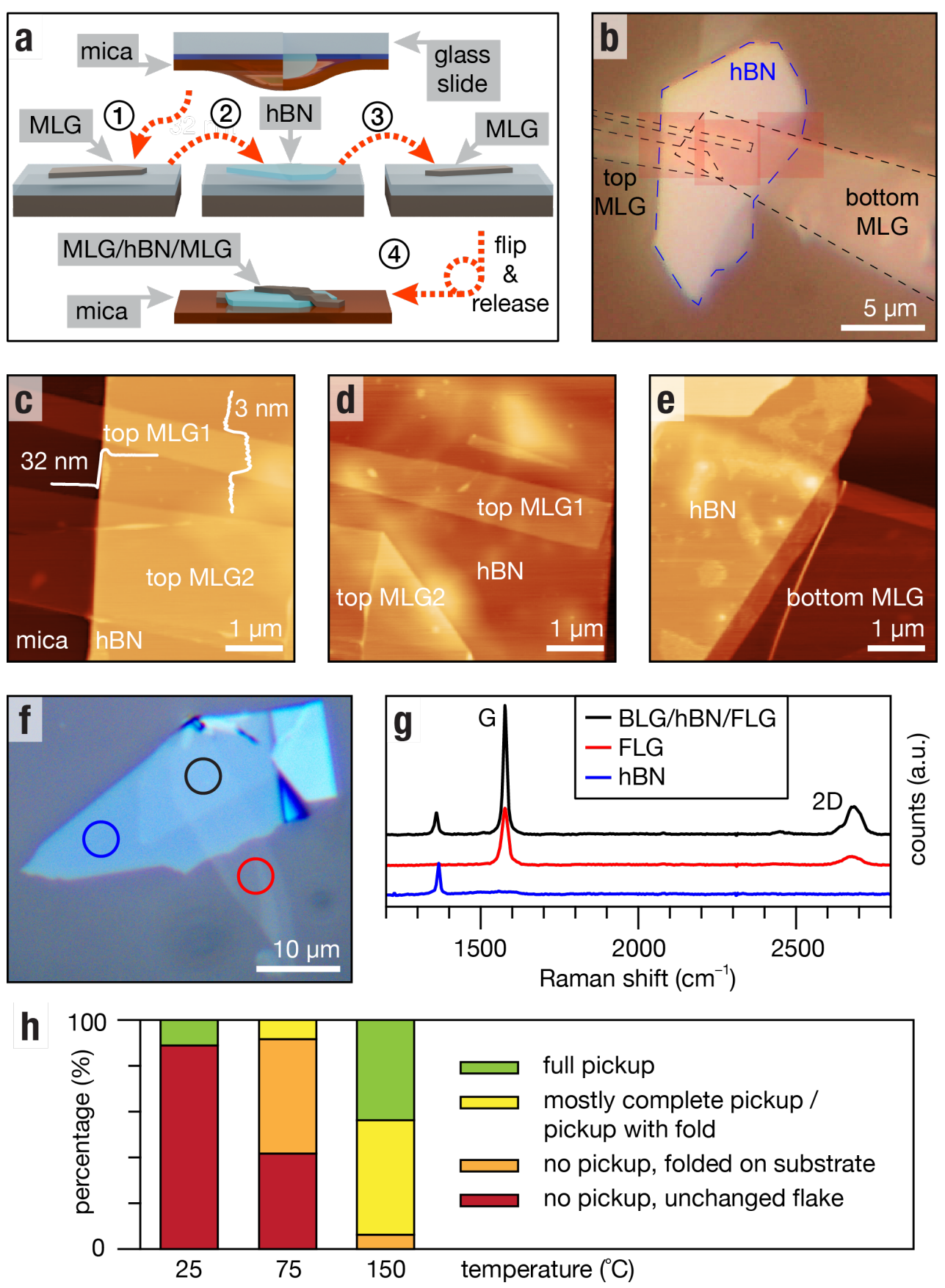

**Figure 2.** Assembly of multilayered vdW heterostructures using mica stamps. (a) Schematic representation of the sequential pick-up transfer steps used in the assembly of a MLG/hBN/MLG stack using a mica stamp. (b) Optical micrograph of the MLG/hBN/MLG/mica stack, (MLG flakes outlined with black dashed lines; hBN flake outlined in blue; red squares indicate the areas of the AFM images in c–e. (c–e) AFM images of the MLG/hBN/MLG stack, showing highly uniform surfaces. (f) Optical micrograph of a BLG/hBN/FLG stack. Circles indicate the positions where Raman spectra were collected. (g) Raman spectra collected at the positions circled in f. (h) Statistic

analysis shows temperature-dependent yield for the pickup of FLG with mica. 9, 12 and 16 trials were performed at 25, 75 and 150 ˚C respectively.

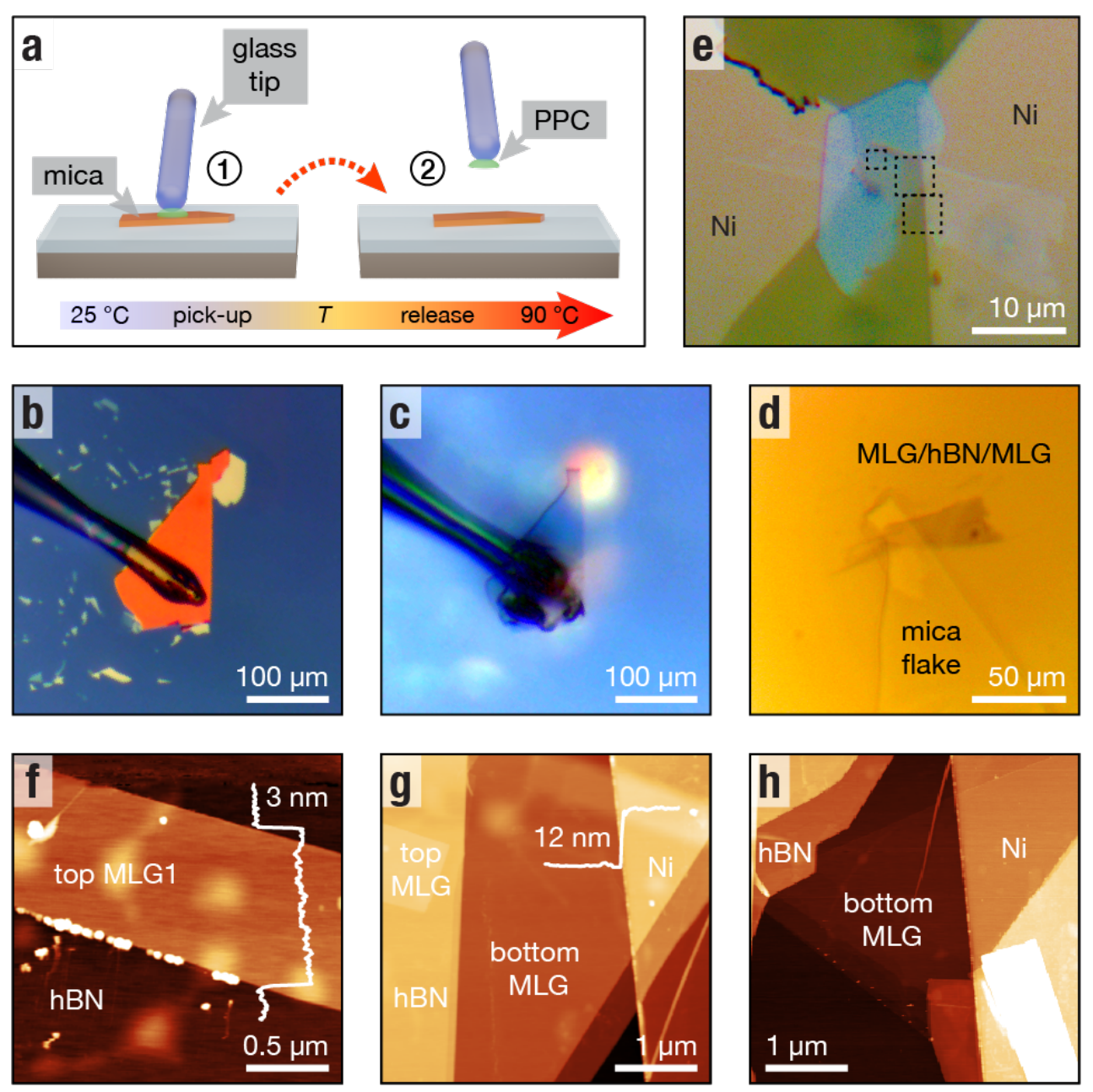

**Figure 3.** Patterning vdW heterostructure contacts using mica shadow masks. (a) Schematic representation of the pick-up and release of mica flakes using a PPC-tipped glass needle. (b) Optical micrograph of a PPC-tipped glass needle contacting a large mica flake (orange) on $SiO_2$. (c) Optical micrograph of the same mica flake adhered to the glass needle, suspended above its original substrate. (d) Optical micrograph of the same mica flake released onto the MLG/hBN/MLG/mica stack from Figure 2. (e) Optical micrograph of the MLG/hBN/MLG/mica with Ni contacts deposited and mica mask removed. Squares outlined with black dashed lines indicate the areas of the following AFM images. (f–h) AFM images of the stack after contact deposition and mask removal, showing highly uniform surfaces.

ASSOCIATED CONTENT

**Supporting Information**. The Supporting Information is available free of charge on the ACS Publications website.

STM of $MoS_2$ transferred onto Au(111), micrographs for assembly of vdW heterostructures, temperature dependence of pickup, photographs of experimental setup, detailed procedure for masking, electrical characterization of a 3-terminal device, AFM characterization of metal contacts. (PDF)

AUTHOR INFORMATION

**Corresponding Author**

* Email: ffischer@berkeley.edu

**Author Contributions**

The manuscript was written through contributions of all authors. All authors have given approval to the final version of the manuscript.

**Funding Sources**

This research program was generously supported by the Heising-Simons Faculty Fellows Program at UC Berkeley.

**Notes**

The authors declare no competing financial interest.

ACKNOWLEDGMENT

This research program was generously supported by the Heising-Simons Faculty Fellows Program at UC Berkeley. H.X.W. acknowledges the Agency for Science, Technology, and Research (A*STAR) Singapore for support through the National Science Scholarship. Metal deposition was done at the UC Berkeley Marvell Nanofabrication Laboratory. We thank the groups of Prof. Michael Crommie and Prof. Kwabena Bediako for discussions and for sharing hBN obtained from the National Institute of Materials Science in Japan.

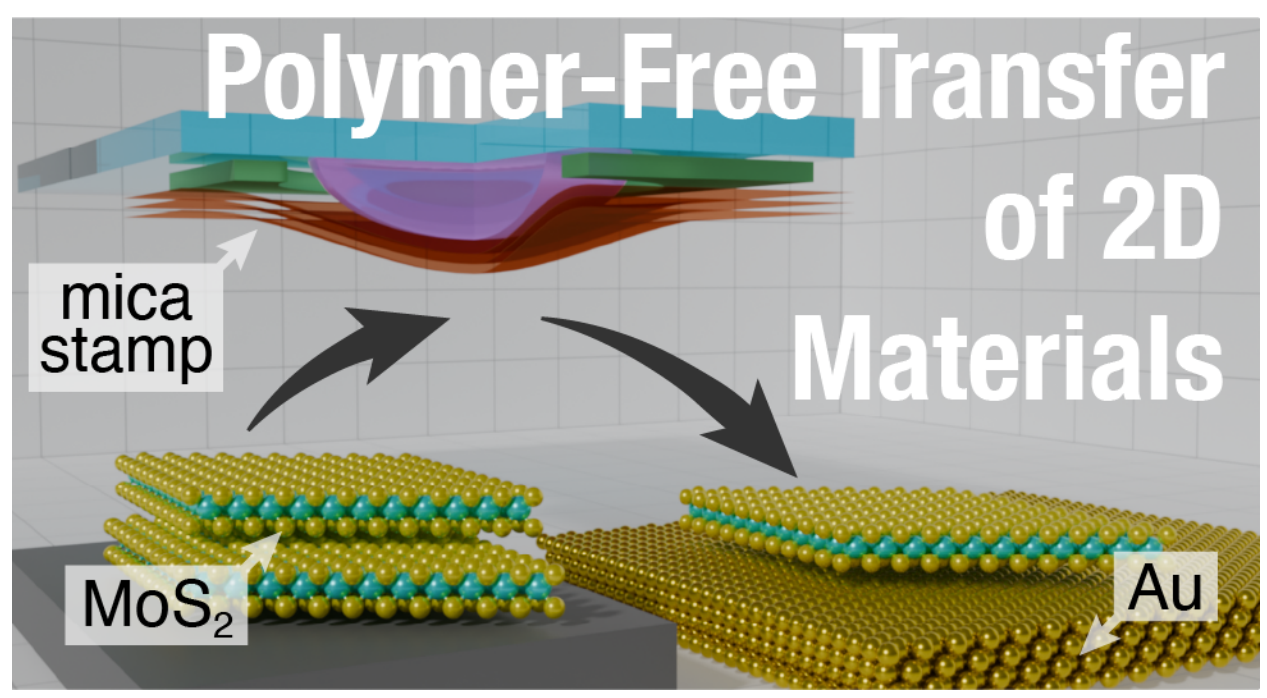